\documentclass[prl,aps,twocolumn,superscriptaddress,amsmath,amssymb]{revtex4}
\usepackage{amsmath}
\usepackage[english]{babel}
\usepackage{graphicx}
\usepackage{bm}
\usepackage{color}

\newcommand{\ds}{\displaystyle}

\begin{document}

\title{ac Hall Effect and Photon Drag of Superconducting Condensate}

\author{S. V. Mironov}
\affiliation{Institute for Physics of Microstructures, Russian Academy of Sciences, 603950 Nizhny Novgorod, GSP-105, Russia}
\author{A. S. Mel'nikov}
\affiliation{Institute for Physics of Microstructures, Russian Academy of Sciences, 603950 Nizhny Novgorod, GSP-105, Russia}
\affiliation{Moscow Institute of Physics and Technology (National Research University), Dolgoprudnyi, Moscow region, 141701 Russia}
\author{A. I. Buzdin}
\affiliation{University Bordeaux, LOMA UMR-CNRS 5798, F-33405 Talence Cedex, France}
\affiliation{World-Class Research Center ``Digital Biodesign and Personalized Healthcare'', Sechenov First Moscow State Medical University, Moscow, 19991, Russia}

\begin{abstract}
We suggest a theoretical description of the photogalvanic phenomena arising in superconducting condensates in the field of electromagnetic wave. The ac Hall effect and photon drag are shown to originate from the second-order nonlinear response of superconducting carriers caused by the suppression of their concentration due to the combined influence of the electron - hole asymmetry and charge imbalance generated by the incident electromagnetic wave. Starting from the time-dependent Ginzburg-Landau theory with the complex relaxation constant we develop a phenomenological description of these phenomena and investigate the resulting behavior of the dc supercurrent and second harmonic induced by microwave radiation incident on a superconductor surface.
\end{abstract}

\maketitle

The ac Hall effect and related photon drag phenomena are intensively studied in various conducting materials starting from the seminal paper by H.E.M. Barlow in 1958 \cite{Barlow}. The basic idea underlying the mechanism of the momentum transfer from photons to the electrons in a solid can be elucidated by the following qualitative arguments. The charge current is known to contain the Hall term proportional to the vector product ${\bf E}\times{\bf B}$ of the electric (${\bf E}$) and magnetic (${\bf B}$) fields in the electromagnetic wave. The photon induced dc current is determined by the time average of this quantity while the time-dependent term gives the second harmonic in the electromagnetic response. These issues have been deeply investigated both theoretically and experimentally for normal metal films and low-dimensional electron systems \cite{Vengurlekar, Strait, Graf, Wieck, Shalygin, Sipe, Grinberg, Gurevich_1, Gurevich_2, Ilyakov, Durnev_1, Durnev_2} as well as for semiconducting systems \cite{Danishevskii, Gibson, Normantas, Mikheev} including exotic compounds like graphene \cite{Karch, Jiang, Glazov}, etc. However, surprisingly the photon drag effect in superconducting systems has not been investigated so far, although it is interesting both from the fundamental standpoint and in view of possible applications in superconducting electronics where it may provide the ultra-fast and energy efficient way to control the dissipationless electric currents with electromagnetic radiation.

The goal of our work is to uncover the origin of these phenomena for the case of superconducting condensates.
At first sight the mechanism of the photon drag of superconducting electrons seems to be more or less obvious if we just adopt a standard Drude-like theory with the infinite momentum relaxation time $\tau_p$ which guarantees us the absence of dissipation. This line of reasoning can be very similar to the one suggested, e.g., in the paper \cite{Ivchenko} for an arbitrary $\tau_p$. However, this type of dynamics of the condensate velocity can be justified only provided the Cooper pair wave function satisfies the Galilean invariant dynamic equation which would disregard completely the superconducting order parameter relaxation \cite{Ivchenko}. Even the simplest microscopic analysis in the gapless limit resulting in the time-dependent Ginzburg-Landau (TDGL) theory \cite{Kopnin_book, Larkin_book} immediately shows that the order parameter relaxation can not be neglected destroying, thus, the above arguments based on the Drude-type consideration. Still, it is known that the relaxation constant in the time-dependent Ginzburg-Landau equation is not exactly real due to the small electron - hole asymmetry of the quasiparticle spectrum \cite{Ebisawa, Kopnin_book, Larkin_book}. It is this asymmetry which is responsible for the partial restoration of the Galilean invariance and, thus, can provide the mechanism of the photon drag of the condensate. Note that the presence of the imaginary part of the relaxation constant follows directly from the gauge invariance of the Ginzburg-Landau theory (see, e.g., \cite{Aronov_1, Aronov_2}). It is also important to emphasize that the imaginary part of the relaxation constant is responsible for the vortex Hall effect \cite{Dorsey, Kopnin_Hall} in the superconducting state and for the inverse Faraday effect \cite{Mironov_IFE, Croitoru_1, Croitoru_2, Croitoru_3} which indicates some general origin of all these phenomena. The importance of the complex relaxation constant has been also noted in previous studies of the photon drag effect in the fluctuating regime above the superconducting critical temperature $T_c$ \cite{Boev_1, Parafilo, Boev_2, Plastovets}. An alternative mechanism of the photon drag phenomenon caused by small changes in the superfluid density due to the electric field effect has been considered in \cite{Radkevich}.

In this manuscript we focus on the study of the ac Hall effect and photon drag below $T_c$ for the well developed superconducting state. For the sake of simplicity we choose the TDGL theory and apply it for two exemplary geometries: superconducting half-space and thin superconducting layer. Accounting the complex valued relaxation constant in TDGL model we get a possibility to describe the second-order nonlinear effects in the microwave response of the condensate responsible for the ac Hall effect and the photon drag phenomena. The mechanism of this second-order nonlinearity is associated with the modulation of the condensate density by the potential of the electron-hole imbalance induced by the incident electromagnetic wave.  This branch imbalance potential is known to play an important role in various problems of nonequilibrium superconductivity (see, e.g., \cite{Artemenko_rev, Tinkham_book} for review) and its relevance to the photoelectric effects has been previously noted in \cite{Aronov, Zaitsev}. 

\begin{figure}[t!]
\begin{center}
\includegraphics[width=0.85\linewidth]{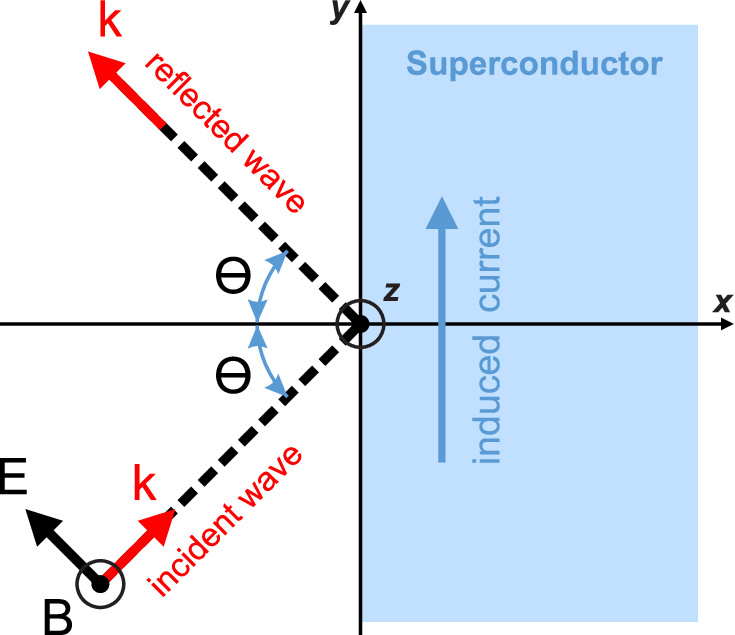}
\end{center}
\caption{Sketch of superconducting half-space irradiated by the electromagnetic wave of E-polarization (magnetic field is parallel to the sample surface).}\label{Fig_System}
\end{figure}

Let us first consider the superconducting (S) sample occupying the half-space $x>0$ under the effect of the electromagnetic wave which has E-polarization (see Fig.~\ref{Fig_System}). The magnetic field in the region $x<0$ reads
\begin{equation}\label{wave_B_field}
B_z = B_0 e^{-i\omega t+ik_y y}\left(e^{ik_x x}+R e^{-ik_x x}\right),
\end{equation}
where $k=\omega/c$ is the absolute value of the wave vector ${\bf k}$, $k_x=k\cos\theta$ and $k_y=k\sin\theta$ are two components of the wave-vector, $B_0$ is the amplitude of the incident wave, and $R$ is the complex reflection coefficient. 

To find the electromagnetic field profile inside the superconductor we adopt the simplest model describing the dynamics of the superconducting order parameter $\psi$. This model is based on the modified TDGL equation with the complex relaxation constant which reflects the electron-hole asymmetry:
\begin{equation}\label{GL_general}
\begin{array}{l}{\ds
\left(\frac{\pi \alpha}{8}+i\gamma\right)\left(\hbar\frac{\partial \psi}{\partial t}-2ie\phi \psi\right)-\alpha T_c\epsilon\psi+}\\{\ds +\alpha T_c\xi_0^2\left(-i\nabla+\frac{2\pi}{\Phi_0}{\bf A}\right)^2\psi+b\left|\psi\right|^2\psi=0,}
\end{array}
\end{equation}
where $\alpha$ and $\gamma$ are the real and imaginary parts of the TDGL relaxation constant, $\phi$ is the electrochemical potential, ${\bf A}$ is the vector potential, $\Phi_0=\pi\hbar c/e$ ($e>0$) is the magnetic flux quantum, $\xi_0$ is the superconducting zero-temperature coherence length, $b$ is the usual Ginzburg-Landau constant, and $\epsilon=1-T/T_c$. Taking the order parameter in the form $\psi=\Delta e^{i\chi}$ we find:
\begin{equation}\label{GL_gen_re}
\frac{\pi \alpha}{8}\hbar\frac{\partial \Delta}{\partial t}+2\gamma e\phi_s \Delta-\left(\alpha T_c\epsilon-m{\bf v}_s^2\right)\Delta -\frac{\hbar^2}{4m}\nabla^2\Delta +b\Delta^3=0,
\end{equation}
\begin{equation}\label{GL_gen_imag}\gamma\frac{\partial \Delta}{\partial t}-\frac{\pi \alpha}{8}\frac{2e}{\hbar}\phi_s\Delta -\frac{1}{2\Delta}{\rm div}\left(\Delta^2{\bf v}_s\right) =0,
\end{equation}
where $m=\hbar^2/(4\alpha T_c\xi_0^2)$ and  
\begin{equation}\label{vs_def}
\phi_s=\phi-\frac{\hbar}{2e}\frac{\partial\chi}{\partial t},~~~{\bf v}_s=\frac{\hbar}{2m}\left(\nabla\chi+\frac{2\pi}{\Phi_0}{\bf A}\right).
\end{equation}
The value $\phi_s$ is a charge imbalance potential proportional to the difference between the chemical potential of normal quasiparticles in superconductor and the chemical potential of Cooper pairs $\mu_p=(\hbar/2e)\partial\chi/\partial t$ \cite{Artemenko_rev}. 

The electric current ${\bf j}={\bf j}_n+{\bf j}_s$ flowing inside the superconductor contains two contributions coming from the normal (${\bf j}_n$) and superconducting (${\bf j}_s$) charge carriers:
\begin{equation}\label{jn_def}
{\bf j}_n=\sigma{\bf E},~~~{\rm where}~~~{\bf E}=-\nabla\phi-\frac{1}{c}\frac{\partial {\bf A}}{\partial t},
\end{equation}
\begin{equation}\label{js_def}
{\bf j}_s=-\frac{c}{4\pi\lambda^2}\left({\bf A}+\frac{\Phi_0}{2\pi}\nabla\chi\right)=-2e\Delta^2{\bf v}_s.
\end{equation}
To treat the magnetic field self-consistently one needs also to take into account the Maxwell equations supplemented with the boundary conditions which require the continuity of the tangential field components $E_y$ and $B_z$ at $x=0$. At the same time, the boundary conditions for the TDGL equation (\ref{GL_general}) require $\partial \Delta/\partial x=0$ and $j_{sx}=0$ at $x=0$. 

The above set of equations uncovers the  mechanisms responsible for the ac Hall and photon drag effects. If one neglects the electron-hole asymmetry the superconducting current (\ref{js_def}) contains only the usual linear response contribution $\propto v_s$ as well as the nonlinear third-order terms $\propto v_s^3$ coming from the corrections $\propto v_s^2$ to the superconducting gap [see Eq.~(\ref{GL_gen_re})]. However, the electron-hole asymmetry gives rise to the oscillatory contribution to $\Delta$ which is determined by the scalar potential $\phi_s$. This contribution results in the generation of both dc and second-harmonic components of the superconducting current which are proportional to $B_0^2$. Below we provide the detailed description of the calculation procedure which results in the final expressions (\ref{Current_0_res_2})-(\ref{Current_2_res_2}).

For the further calculations it is convenient to introduce the complex amplitudes $\tilde u$ for all oscillating variables $u={\rm Re}\left(\tilde ue^{-i\omega t}\right)$ and choose $\tilde u(x,y)\propto e^{ik_y y}$ due to the translational symmetry of the system and the boundary conditions. To calculate the induced superconducting current we use a perturbation theory with respect to the small parameter $\gamma/\alpha$ and the incident wave amplitude $B_0$. Within this strategy we first neglect the deviation of the gap function from the equilibrium value $\Delta_0=\sqrt{(\alpha T_c/b)\epsilon}$ and take the London penetration depth $\lambda=\sqrt{mc^2/\left(8\pi e^2\Delta_0^2 \right)}$ to be constant. Thus, in complex amplitudes the expression for the supercurrent reads:
\begin{equation}\label{js2}
\tilde{\bf j}_s=\frac{ic^2}{4\pi \omega\lambda^2}\left(\tilde{\bf E}+\nabla\tilde\phi_s\right).
\end{equation}
Then from the Maxwell equation we get:
\begin{equation}\label{Max1}
{\rm curl}~\tilde{\bf B}=\left(\frac{4\pi\sigma}{c}+\frac{ic}{\omega\lambda^2}-\frac{i\omega}{c}\right)\tilde{\bf E}+\frac{ic}{\omega\lambda^2}\nabla\tilde\phi_s.
\end{equation}
Taking the curl of Eq.~(\ref{Max1}) and accounting the Maxwell equations we find the equation for the field ${\bf B}$:
\begin{equation}\label{Max2}
\nabla^2\tilde{\bf B}=\frac{1}{\lambda_{\rm eff}^2}\tilde{\bf B},~~~{\rm where}~~~\frac{1}{\lambda_{\rm eff}^2}= \frac{1}{\lambda^2}-\frac{4\pi i\sigma\omega}{c^2}-\frac{\omega^2}{c^2}.
\end{equation}
For simplicity we consider the limit of low $\omega$ values so that the wave-length is much larger than both the skin-layer depth $\delta=c/\sqrt{4\pi\sigma\omega}$ and the London penetration depth, i.e. $\left(\omega/c\right)^2\ll {\rm min}\left\{4\pi\sigma\omega/c^2;~\lambda^{-2}\right\}$. Then we may neglect the last term in the expression for $\lambda_{\rm eff}^{-2}$ as well as the derivative $\partial^2\tilde{\bf B}/\partial y^2\propto (\omega/c)^2\tilde{\bf B}$  in Eq.~(\ref{Max2}). The resulting profile for the only nonzero magnetic field component inside the superconductor reads $\tilde B_z(x,y)=\tilde B_z^{(0)}\exp\left(ik_y y-x/\lambda_{\rm eff}\right)$, where $\tilde B_z^{(0)}$ is the complex constant determined by the boundary conditions. 

At the same time, the combination of Eq.~(\ref{Max1}),  TDGL equation (\ref{GL_gen_imag})  and Eq.~(\ref{js2}) allows one to obtain the equation for the potential $\tilde\phi_s$. To do this we may rewrite Eq.~(\ref{GL_gen_imag}) using Eqs.~(\ref{js_def}) and (\ref{js2}) and also neglect the term $\gamma(\partial\Delta/\partial t)$ which has the order of $(\gamma/\alpha)^2$: 
\begin{equation}\label{GL_gen_imag2}
-\frac{\pi \alpha}{8}\frac{2e}{\hbar}\tilde\phi_s\Delta_0^2 +\frac{ic^2}{14\pi e \omega\lambda^2}{\rm div}\left(\tilde{\bf E}+\nabla\tilde\phi_s\right) =0.
\end{equation}
Then taking the divergence of Eq.~(\ref{Max1}) we find that ${\rm div}~\tilde{\bf E}=-\left(\lambda_{\rm eff}/\lambda\right)^2\nabla^2\tilde\phi_s$, and Eq.~(\ref{GL_gen_imag2}) becomes
\begin{equation}\label{GL_gen_imag3}
l_E^2\frac{\lambda_{\rm eff}^2}{\lambda^2}\nabla^2\tilde\phi_s =\tilde\phi_s,
\end{equation}
where $l_E=\sqrt{\hbar\sigma/(\pi e^2\alpha\Delta_0^2)}$ is the length describing the scale of conversion between normal and superconducting currents. For simplicity we consider the limit of well developed superconductivity so that $(\omega/c)\ll\l_E^{-1}$ and the derivative $(\partial^2\tilde\phi_s/\partial y^2)$ in Eq.~(\ref{GL_gen_imag3}) can be neglected. Then one finds
\begin{equation}\label{phis_res}
\tilde\phi_s(x,y)=\tilde\phi_s^{(0)}\exp\left(ik_y y-\frac{\lambda}{\lambda_{\rm eff}}\frac{x}{l_E}\right),
\end{equation}
where $\tilde\phi_s^{(0)}$ is a certain complex constant.

As a next step, we use the boundary conditions for the Maxwell and TDGL equations to find the constants $\tilde B_z^{(0)}$ and $\tilde\phi_s^{(0)}$. A standard boundary condition for the order parameter gives us the absence of the supercurrent through the superconductor surface ($j_{sx} = 0$) and resulting relation $\tilde E_x(0)=-\left.\left(\partial \tilde\phi_s/\partial x\right)\right|_{x=0}$. Then from the $x$-component of Eq.~(\ref{Max1}) we obtain 
\begin{equation}\label{Max3b_gen}
\left.\frac{\partial\tilde\phi_s}{\partial x}\right|_{x=0}=-\frac{i\omega}{4\pi\sigma}\sin\theta \tilde B_z^{(0)},
\end{equation}
which gives [see Eq.~(\ref{phis_res})]
\begin{equation}\label{Max3b}
\tilde\phi_s^{(0)}=\frac{i\omega}{4\pi\sigma}\frac{\lambda_{\rm eff}l_E}{\lambda}\sin\theta \tilde B_z^{(0)}.
\end{equation}
At the same time, the continuity of the field components $B_z$ and $E_y$ at the boundary $x=0$ provides two other equations: $\tilde B_z^{(0)}=(1+R)B_0$ and
\begin{equation}\label{BC_2}
\frac{i\omega}{c}\left[\lambda_{\rm eff}\tilde B_z^{(0)}-\frac{\lambda_{\rm eff}^2}{\lambda^2}\sin\theta\tilde\phi_s^{(0)}\right]=(1-R)\cos\theta B_0.
\end{equation}
In the low-frequency limit $\omega\ll 4\pi\sigma$ the expression in the left-hand side of Eq.~(\ref{BC_2}) is much smaller than $B_0\cos\theta$ (at least, for not very large angles $\theta$) which corresponds to $R\approx 1$ and $\tilde B_z^{(0)}\approx 2B_0$. The expression for $\tilde\phi_s^{(0)}$ then directly follows from Eq.~(\ref{Max3b}).

Note that when dealing with the electromagnetic waves irradiating conductive media one often uses the so-called Leontovich boundary condition which assumes that inside the media the electric and magnetic field components are almost parallel to the surface (see, e.g., \cite{Landau_book}). Obviously to describe the generation of the charge imbalance potential inside the sample and the subsequent generation of the dc and second harmonic response we should go beyond this approximation and take into account the electric field component $E_x$ perpendicular to the surface.

Substituting the obtained expressions for the potential $\tilde\phi_s$ into the Ginzburg-Landau equation and using the perturbative approach we search the solution of Eq.~(\ref{GL_gen_re}) in the form $\Delta=\Delta_0+\tilde\Delta_1(x)e^{-i\omega t+ik_yy}$, where the correction $\tilde\Delta_1$ is small (see the calculation details in \cite{supp}). After that considering the low-frequency limit $\omega/c\ll\xi^{-1}$ (here $\xi=\xi_0/\sqrt{\epsilon}$ is the temperature-dependent superconducting correlation length) we use the obtained expressions for $\tilde\Delta_1$, $\tilde {\bf E}$ and $\tilde\phi_s$ to find the value $\tilde {\bf v}_s=-(ie/m\omega)\left(\tilde {\bf E}+\nabla\tilde\phi_s\right)$ and calculate the superconducting current flowing along the sample boundary. Within our perturbative approach it contains three contributions:
\begin{equation}\label{curr_def}
j_{sy}=j_{sy}^{(0)}+j_{sy}^{(\omega)}+j_{sy}^{(2\omega)}.
\end{equation}
The first one $j_{sy}^{(0)}=-2e\Delta_0{\rm Re}\left\{\tilde\Delta_1 \tilde v_{sy}^*\right\}$ is the dc component of the current responsible for the photon drag effect, the second one $j_{sy}^{(\omega)}=-2e\Delta_0^2{\rm Re}\left\{\tilde v_{sy}e^{-i\omega t}\right\}$ is the usual linear response contribution while the third one $j_{sy}^{(2\omega)}=-2e\Delta_0{\rm Re}\left\{\tilde\Delta_1 \tilde v_{sy}e^{-2i\omega t}\right\}$ is the second harmonic response which is proportional to $B_0^2$ and oscillates at $2\omega$ frequency. In what follows, we will not be interested in the spatial distribution of the superconducting current near the sample boundary and calculate only the total surface current density $I_y=\int\limits_0^\infty j_{sy}(x)dx$. Also, for simplicity we will focus on type-II superconductors where the lengths $\xi$ and $l_E$ are much smaller than $\lambda$. 

\begin{figure}[t!]
\begin{center}
\includegraphics[width=0.95\linewidth]{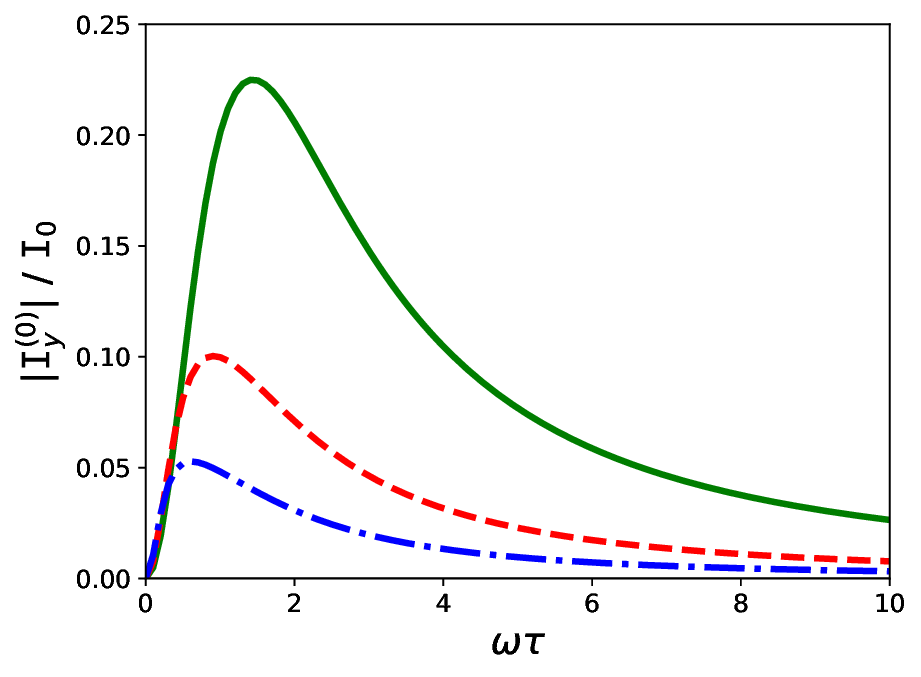}
\end{center}
\caption{Dependence of the d.c. current arising due to photon drag effect on the radiation frequency. The green solid, red dashed and blue dash-dotted curves correspond to the following values of the parameter $l_E/\xi$: $1.0$, $1.5$ and $2.0$. The current is normalized by the value $I_0=\left(\nu/\sqrt{2}\pi\right)\left(\xi/\tau\right)\left( B_0^2\sin\theta/H_{cm}\right)$.}\label{Fig_res}
\end{figure}

Both the zero-frequency response $I_y^{(0)}$ and the response on the second harmonic $I_y^{(2\omega)}$ strongly depend on the radiation frequency. Taking into account that $\lambda^2/\lambda_{\rm eff}^2=1-i\omega\tau\eta$ (here $\eta=l_E^2/\xi^2$) we find the d.c. current
\begin{equation}\label{Current_0_res_2}
I_y^{(0)}= \frac{\nu}{\sqrt{2}\pi}\frac{\xi}{\tau}\frac{B_0^2\sin\theta}{H_{cm}} \frac{\omega\tau}{\left(1+\omega^2\tau^2\eta^2\right)}{\rm Re}\left[\frac{i\sqrt{1+i\omega\tau\eta}}{\left(2-i\omega\tau\right)}\right],
\end{equation}
where the parameter $\nu=8\gamma/\pi\alpha$ characterizes the ratio between the imaginary and real parts of the TDGL relaxation constant, $\tau=\pi\hbar/(8T_c\epsilon)$ is the characteristic time-scale of the TDGL theory, and $H_{cm}= \Phi_0/(2\sqrt{2}\pi\xi\lambda)$ is the thermodynamic critical field. Note that for clean superconductors $\xi_0\sim \hbar v_F/\Delta_0$ (where $v_F$ is the Fermi velocity) so that $\xi/\tau\sim v_F\sqrt{1-T/T_c}$. The dependencies of  $I_y^{(0)}$ on the parameter $\omega\tau$ for different ratios $\l_E/\xi$ are shown in Fig.~\ref{Fig_res}. In the small-frequency limit the d.c. current is proportional to $\omega^2$ while in the high-frequency limit it decays as $I_y^{(0)}\propto\omega^{-3/2}$. Similarly, for the response on the second harmonic we obtain:
\begin{equation}\label{Current_2_res_2}
I_y^{(2\omega)}= \frac{\nu}{\sqrt{2}\pi}\frac{\xi}{\tau}\frac{B_0^2\sin\theta}{H_{cm}} {\rm Re}\left[ \frac{i\omega\tau e^{2ik_yy-2i\omega t}}{\left(2-i\omega\tau\right)\left(1-i\omega\tau\eta\right)^{3/2}}\right].
\end{equation}

In the above calculations we assumed the thickness of the superconducting sample to be larger than all other characteristic length scales relevant to the problem. However, the described generic formalism is also suitable for the description of photogalvanic effects in superconducting samples of arbitrary thickness. For illustration we consider the opposite limiting case of thin superconducting film of the thickness $d_s\ll c/(4\pi\sigma)$ occupying the region $0< x < d_s$. Because of the small film thickness the amplitude of reflected wave is negligibly small so that the magnetic field outside the superconducting film reads $B_z=B_0 e^{-i\omega t+ik_x x+ik_y y}$.

Considering the limit of small $d_s$ we  solve Eq.~(\ref{GL_gen_imag3}) together with Eq.~(\ref{Max3b_gen}), then analyze the time dependence of the correction $\tilde\Delta_1$ to the superconducting gap and finally calculate the superconducting current (see the calculation details in \cite{supp}). Assuming for simplicity that $k_y\xi\ll\omega\tau\ll 1$ and $(l_E^2/\lambda^2)(\omega/4\pi\sigma)\ll 1$ for the dc component of the current we obtain:
\begin{equation}\label{Film_current_dc}
I_{y}^{(0)}=\frac{\nu c^2\xi^2 B_0^2\tau d_s\cos^2\theta\sin\theta}{2\lambda^2\Phi_0\left(4+\omega^2\tau^2\right)},
\end{equation}
while the expression for the second harmonic response reads $I_{y}^{(2\omega)}=I_{y}^{(0)}\left(\omega\tau\right)^{-1}{\rm Re}\left[\left(2i-\omega\tau\right)e^{-2i\omega\tau+2ik_yy}\right]$ where $I_{y}^{(0)}$ is defined by Eq.~(\ref{Film_current_dc}).
These expressions substantially differ from the ones obtained for the superconducting half-space: (i) there appears an additional $\cos^2\theta$ factor caused by negligibly small reflection of the incident wave in the film geometry; (ii) the typical scales of the field and potentials decay in the results for the half-infinite superconductor should be replaced by the film thickness. 

Note that exactly the same result for the current induced in atomically thin superconducting film can be obtained by solving the TDGL equation with the vector potential $A_y=-(ik_x/k^2)B_0e^{-i\omega t+ik_x x+ik_yy}$ corresponding to the incident electromagnetic wave. In this case the dynamics of the gap potential $\Delta$ arises due to nonzero derivative $\partial A_y/\partial y\neq 0$ which enters the TDGL equation and gives rise to the oscillations of $\Delta$. Interestingly, there exists a formal equivalence between the above problem and the inverse Faraday effect in the superconducting ring of the radius $R$ irradiated by the circularly polarized light \cite{Croitoru_2}.  The expression (\ref{Film_current_dc}) for the dc current can be directly obtained from the result of Ref.~\cite{Croitoru_2}  if one formally replaces the combination $R\vartheta$ (where $\vartheta$ is the azimuthal angle coordinate) in the equations describing the ring geometry by the coordinate $y$ relevant for the geometry studied in the present paper (see the details in \cite{supp}).

To estimate the possible values of the  photon drag effect for typical low-temperature superconductors we consider, e.g., the Nb film with $T_c\sim10~{\rm K}$, $\nu\sim 10^{-3}$, $\xi\sim 30~{\rm nm}$ and $\lambda\sim 100~{\rm nm}$ irradiated by the wave of the frequency $\omega\sim 10^{12}~{\rm sec}^{-1}$ and intensity ${\mathcal I}=cB_0^2/8\pi\sim 10~ {\rm \mu W}/{\rm \mu m}^2$. Taking the temperature slightly below $T_c$ so that $\epsilon\sim 0.1$ and $\theta=\pi/4$ we get the dc current density $j_{sy}^{(0)}\sim 10^2~{\rm A}/{\rm cm}^2$. This value is comparable with the magnitude of the photon drag effect in normal metals.  However, the contribution from the superconducting condensate to the photo-induced dc current strongly depends on temperature near $T_c$ which should allow its separation from the one coming from normal electrons. It is natural to expect that superconducting films of FeSe where $T_c$ is comparable with the Fermi energy $E_F$ \cite{Kasahara} or high-temperature cuprates with $T_c/E_F\sim 0.1$ \cite{Blatter} may provide higher magnitude of the induced currents. However, the increase in the factor $\nu\sim T_c/E_F$ appears to be counter-balanced by the relatively small values of the ratio $\xi/\lambda\sim 10^{-2}$ in these materials and the estimates give $j_{sy}^{(0)}\sim 10^2~{\rm A}/{\rm cm}^2$ for the films of FeSe (which is comparable with the effect in Nb) and $j_{sy}^{(0)}\sim 10~{\rm A}/{\rm cm}^2$ for cuprates.

To observe the predicted dc current experimentally one may use, e.g., the superconducting sample embedded into a conducting loop (similar geometry was considered in Ref.~\cite{Gurevich_1}). In this case the generation of the optically controlled current should be accompanied by the changes of the magnetic flux trapped inside the loop which can be detected in the magnetic field measurements. If the circuit is broken the radiation induces a nonzero phase difference between the opposite ends of the superconductor (similarly to the so-called phase batteries based on the anomalous Josephson effect \cite{Buzdin_phi, Goldobin_1, Goldobin_2}). At the same time, the second harmonic response can be detected by analyzing the spectrum of the radiation  from the sample.  

Another peculiar feature of the predicted superconducting photon drag effect as well as the second-harmonic response is the dependence of the current direction on the sign of the value $\nu$, i.e. the sign of the imaginary part $\gamma$ of the TDGL relaxation constant. This imaginary part it known to be responsible for the sign change in the Hall effect in superconductors \cite{Artemenko, Dorsey, Kopnin_Hall} and can take both positive and negative values. This means that the directions of the photo-induced currents associated with normal and superconducting electrons can have opposite direction and, in principle, even cause the reversal of the total current direction in the vicinity of the superconducting critical temperature.   

Note finally that the phenomenon of the modulation of the superconducting order parameter by the branch imbalance potential oscillations induced by the incident electromagnetic wave  provides a hint towards a novel mechanism of the Higgs mode generation in superconductors. The above TDGL model can not, of course, describe the regime of gapped superconductivity and related Higgs-type gap oscillations \cite{Pekker}. However, we expect that accounting the electron-hole asymmetry effects within more elaborated microscopic theory (see, e.g., \cite{Volkov_Kogan, Yuzbashyan}) one can obtain a resonant excitation of superconducting gap oscillations, i.e. the Higgs modes,  which can be observed through the measurements of the drag effect and second-order harmonic response.

\begin{acknowledgments}

The authors thank S. A. Tarasenko for stimulating discussions. This work was supported by the Russian Science Foundation (Grant No. 21-72-10161) in part related to the solution of a half-space problem and by the Ministry of Science and Higher Education of the Russian Federation (contract No. 075-15-2022-316 of the Center of Excellence ``Center of Photonics'' and project No. FSMG-2023-0011), in part related to a thin-film problem. S. V. M. acknowledges the financial support of the Foundation for the Advancement of Theoretical Physics and Mathematics BASIS (Grant No. 23-1-2-32-1). The work of A. I. B. was supported by ANR SUPERFAST and the LIGHT S\&T Graduate Program and EU COST CA21144  Superqumap. 

\end{acknowledgments}

\renewcommand{\theequation}{S\arabic{equation}}

\setcounter{equation}{0}

\section*{Details of the perturbative solution of the Ginzburg-Landau equation}

To analyze the radiation-stimulated dynamics of the gap function $\Delta$ in the superconducting half-space we assume the light-induced effects to be small and search the solution of the Ginzburg-Landau equation in the form $\Delta=\Delta_0+\tilde\Delta_1(x)e^{-i\omega t+ik_yy}$, where the correction $\tilde\Delta_1$ is small. Then linearizing TDGL equation over $\Delta_1$ and the velocity $v_s$ we obtain:
\begin{equation}\label{GL_gen_re2}
-i\omega\tau \tilde\Delta_1+2\tilde\Delta_1 -\xi^2\nabla^2\tilde\Delta_1 =-\frac{\pi\nu}{4} \frac{\Delta_0}{ T_c\epsilon}e\tilde\phi_s ,
\end{equation}
where $\xi=\xi_0/\sqrt{\epsilon}$ is the temperature-dependent superconducting correlation length, $\tau=\pi\hbar/(8T_c\epsilon)$ is the characteristic time-scale of the TDGL theory, and the parameter $\nu=8\gamma/\pi\alpha$ characterizes the ratio between the imaginary and real parts of the TDGL relaxation constant. In what follows we consider the low-frequency limit and assume that $\omega/c\ll\xi^{-1}$. The solution of Eq.~(\ref{GL_gen_re2}) satisfying the boundary condition $\left.\left(\partial\tilde\Delta_1/\partial x\right)\right|_{x=0}=0$ reads
\begin{equation}\label{Delta_sol}
\tilde\Delta_1 = \frac{i\nu\omega e\Delta_0 B_0\sin\theta}{ 8\sigma \xi_0^2T_c}  \frac{e^{ik_yy}}{q_2^2-q_1^2}\left(\frac{1}{q_2}e^{-q_2x}-\frac{1}{q_1}e^{-q_1x}\right),
\end{equation}
where $q_1=\xi^{-1}\left(\sqrt{\sqrt{1+w^2}+1}-i\sqrt{\sqrt{1+w^2}-1}\right)$,  $w=\omega\tau/2$, and $q_2=\lambda/(\lambda_{\rm eff}l_E)$.

\section*{Details of the solution of the TDGL equation in thin superconducting film}

Here we calculate the radiation-induced superconducting current in the thin superconducting film. Assuming $d_s$ to be small we may integrate Eq.~(12) over the film thickness which gives
\begin{equation}\label{GL_gen_integr}
l_E^2\frac{\lambda_{\rm eff}^2}{\lambda^2}\left(d_s\nabla_{yz}^2\tilde\phi_s+\left.\frac{\partial \tilde\phi_s}{\partial x}\right|_{d_s}-\left.\frac{\partial \tilde\phi_s}{\partial x}\right|_{0}\right) =\tilde\phi_sd_s,
\end{equation}
where $\nabla^2_{yz}=\partial^2/\partial y^2+\partial^2/\partial z^2$. The difference between the derivatives $\partial \tilde\phi_s/\partial x$ at the sample boundaries can be expressed through the difference between the corresponding magnetic field values using Eq.~(14) and analogous relation for another boundary of the film. Since the film thickness is small we may put $\tilde B_z(d_s)-\tilde B_z(0)\approx \left(\partial \tilde B_z/\partial x\right)d_s$, then substitute the derivative $\partial \tilde B_z/\partial x$ from Eq.~(9) and after algebraic transformations obtain
\begin{equation}\label{GL_gen_integr2}
-\frac{l_E^2}{\lambda^2}\frac{c}{4\pi\sigma}\sin\theta\left(\tilde E_y+i\frac{\omega}{c}\sin\theta\tilde\phi_s\right)=\tilde\phi_s.
\end{equation}
In the small frequency limit when $(l_E^2/\lambda^2)(\omega/4\pi\sigma)\ll 1$ the second term inside the brackets can be neglected. Then accounting the continuity of the electric field component $\tilde E_y$ at $x=0$ we find that $\tilde E_y=B_0\cos\theta e^{ik_yy}$ which gives $\tilde\phi_s=-\left(l_E/\lambda\right)^2\left(c/4\pi\sigma\right)B_0\sin\theta \cos\theta e^{ik_yy}$. The obtained expression for the potential $\tilde\phi_s$ allows us to calculate the correction to the gap potential $\tilde \Delta_1$ and then obtain the expression (20) for the superconducting.

\section*{Alternative approach for description of the photo-galvalic effects in atomically thin superconducting films}

The results for the dc current and the second-harmonic electromagnetic response for the atomically thin superconducting film can be also obtained by solving the GL equation with the vector potential corresponding to the incident electromagnetic wave. Indeed, considering the film geometry described in the paper and neglecting the wave reflection we may assume that the magnetic field acting on the superconducting condensate inside the film coincides with the field of the incident wave:
\begin{equation}\label{B_field_def}
B_z=B_0 e^{-i\omega t+ik_x x+ik_y y}.
\end{equation}  
The corresponding vector potential at $x=0$ has two components directed parallel and perpendicular to the film surface, the first one being $A_y=-(ik_x/k^2)B_0e^{-i\omega t+ik_x x+ik_y y}$. Note that inside the film (i.e. in the plane $x=0$) the derivative $\partial A_y/\partial y =B_0\cos\theta\sin\theta e^{-i\omega t+ik_y y}\neq 0$.

The time-dependent GL equations take the form:
\begin{equation}\label{GL_Re}
\tau\frac{\partial \Delta_1}{\partial t}-\tau \nu\Delta_0\frac{\partial \chi}{\partial t}+2\Delta_1-\xi^2\nabla^2\Delta_1=0,
\end{equation}
\begin{equation}\label{GL_Im}
\nu\tau\frac{\partial \Delta_1}{\partial t}+\tau\Delta_0\frac{\partial \chi}{\partial t}-\xi^2\Delta_0\nabla^2\chi=\Delta_0\xi^2\frac{2\pi}{\Phi_0}\frac{\partial A_y}{\partial y}.
\end{equation}
Searching the solution in the form $u = {\rm Re}\left(\tilde{u} e^{-i\omega t}\right)$ we may rewrite these equations as follows:
\begin{equation}\label{GL_Re2}
-i\omega\tau\tilde\Delta_1+i\omega\tau \nu\Delta_0\tilde\chi+2\tilde\Delta_1+k_y^2\xi^2\tilde\Delta_1=0,
\end{equation}
\begin{equation}\label{GL_Im2}
-i\nu\omega\tau\tilde\Delta_1-i\omega\tau\Delta_0\tilde\chi+k_y^2\xi^2\Delta_0\tilde\chi=\Delta_0\xi^2\frac{2\pi}{\Phi_0}\frac{B_0k_xk_y}{k^2}.
\end{equation}
Solving these algebraic equations we find:
\begin{equation}\label{Film_chi_res}
\tilde\chi=\frac{2\pi \xi^2 B_0}{\Phi_0}\frac{\cos\theta\sin\theta e^{ik_yy}}{k_y^2\xi^2-i\omega\tau},
\end{equation}
\begin{equation}\label{Film_delta_res}
\tilde\Delta_1=-\frac{2\pi \xi^2 B_0}{\Phi_0}\frac{i\omega\tau \nu\Delta_0\cos\theta\sin\theta e^{ik_yy}}{\left(2+k_y^2\xi^2-i\omega\tau\right)\left(k_y^2\xi^2-i\omega\tau\right)}.
\end{equation}
Then
\begin{equation}
\tilde v_y=-\frac{k_x\omega\tau B_0 e^{ik_yy}}{k^2\left(k_y^2\xi^2-i\omega\tau\right)}.
\end{equation}

Substituting these results to the expression for the superconducting current in the limit of small frequencies when $k_y^2\xi^2 \ll \omega\tau\ll 1$ we obtain:
\begin{equation}\label{Film_current_dc}
j_{sy}^{(0)}=\frac{\nu c^2\xi^2 B_0^2\tau\cos^2\theta\sin\theta}{2\lambda^2\Phi_0\left(4+\omega^2\tau^2\right)},
\end{equation}

\begin{equation}
\label{Film_current_2omega}
\begin{array}{c}{\ds j_{sy}^{(2\omega)}=-\frac{\nu c^2 \xi^2 B_0^2   \cos^2\theta\sin\theta }{ 2\lambda^2\Phi_0 }}\\{}\\{\ds \times\frac{\omega\tau \cos\left(2\omega t-2k_yy\right)-2\sin\left(2\omega t-2k_yy\right)}{\omega\left(4+\omega^2\tau^2\right)}.}
\end{array}
\end{equation}

Interestingly, Eqs.~(\ref{GL_Re})-(\ref{GL_Im}) formally coincide with the ones used in Ref.~\cite{Croitoru_2} for the case of the superconducting ring radiated by the circularly polarized electromagnetic wave. Indeed, let us consider the ring of the radius $R$ positioned in the plane $x=0$ and introduce the azimuthal angle coordinate $\vartheta$ along the ring. The electric field in the plane of the ring can be chosen in the form ${\bf E}=E_0\left({\bf e}_z-i{\bf e}_y\right)e^{-i\omega t}$ with the corresponding vector potential $A_\vartheta=-(icE_0/\omega)e^{i\vartheta-i\omega t}$. Then the the GL equations take the form
\begin{equation}\label{GL_ring_Re}
\tau\frac{\partial \Delta_1}{\partial t}-\tau \nu\Delta_0\frac{\partial \chi}{\partial t}+2\Delta_1-\xi^2\frac{1}{R^2}\frac{\partial^2 \Delta_1}{\partial \vartheta^2}=0,
\end{equation}
\begin{equation}\label{GL_ring_Im}
\nu\tau\frac{\partial \Delta_1}{\partial t}+\tau\Delta_0\frac{\partial \chi}{\partial t}-\xi^2\Delta_0\frac{1}{R^2}\frac{\partial^2 \chi}{\partial \vartheta^2}=\Delta_0\xi^2\frac{2\pi}{\Phi_0}\frac{1}{R}\frac{\partial A_\vartheta}{\partial \vartheta}.
\end{equation}
Using the complex amplitudes we may rewrite this system in the form
\begin{equation}\label{GL_ring_Re2}
-i\omega\tau\tilde\Delta_1+i\omega\tau \nu\Delta_0\tilde\chi+2\tilde\Delta_1+\frac{\xi^2}{R^2}\tilde\Delta_1=0,
\end{equation}
\begin{equation}\label{GL_ring_Im2}
-i\nu\omega\tau\tilde\Delta_1-i\omega\tau\Delta_0\tilde\chi+\Delta_0\frac{\xi^2}{R^2}\tilde\chi=\Delta_0\xi^2\frac{2\pi}{\Phi_0}\frac{1}{R}\frac{cE_0}{\omega}.
\end{equation}
Comparing the systems of equations (\ref{GL_ring_Re2})-(\ref{GL_ring_Im2}) and (\ref{GL_Re2})-(\ref{GL_Im2}) one sees that the formal substitutions $R\to 1/k_y$ and $E_0\to B_0(k_x/k)=B_0\cos\theta$ in Eqs.~(\ref{GL_ring_Re2})-(\ref{GL_ring_Im2}) make these two systems equivalent which allows to use the relations found in \cite{Croitoru_2} to obtain the expression (\ref{Film_current_dc}).

\end{document}